\documentclass[reprint,superscriptaddress,nobibnotes,amsmath,amssymb,twocolumn,prb]{revtex4-2}
\usepackage[pdftex]{graphicx}
\usepackage[english]{babel}
\usepackage{physics}
\usepackage{tabularx}
\usepackage{float}
\usepackage{amssymb}   % for math
\usepackage{amsmath}   % for math
\usepackage{dsfont}   % for math
\usepackage{epsfig}
\usepackage{dcolumn}   % needed for some tables
\usepackage{bm}        % for math
\usepackage{amstext}
\usepackage{epsfig}
\usepackage{mdframed}
\newcommand{\be}{\begin{equation}}
\newcommand{\ee}{\end{equation}}

\newcommand{\LCPQ}{Laboratoire de Chimie et Physique Quantiques (UMR 5626), Universit\'e de Toulouse, CNRS, UPS, France}

\begin{document}
\title{Path integral for the quartic oscillator: An accurate analytic formula 
for the partition function}
\author{Michel Caffarel}
\email[]{caffarel@irsamc.ups-tlse.fr}
\affiliation{\LCPQ}
%%%%%%%%%%%%%%%%%%%%%%%%%%%%%%%%%%%%%%%%%%%%%%%%%%%%%%%%%%%%%%%%%%%%%%%%%%%%%%%%%%%%%%%%%%%%%%%%%%%%%%%%%%%%%%%%%%%%%%%%%%%%%%%
\begin{abstract}

In this work an approximate analytic expression for the quantum partition function of the quartic 
oscillator described by the potential $V(x) = \frac{1}{2} \omega^2 x^2 + g x^4$ is presented.
Using a path integral formalism, the exact partition function is approximated by the partition function 
of a harmonic oscillator with an effective frequency depending both on the temperature and coupling constant $g$.
By invoking a Principle of Minimal Sensitivity (PMS) of the path integral to the effective frequency, 
we derive a mathematically well-defined analytic formula for the partition function. 
Quite remarkably, the formula reproduces qualitatively and quantitatively the key features 
of the exact partition function.
The free energy is accurate to a few percent over the entire range of temperatures and coupling strengths $g$.
Both the harmonic ($g\rightarrow 0$) and classical (high-temperature) limits are exactly recovered. 
The divergence of the power series of the ground-state energy at weak coupling, characterized by a factorial growth of 
the perturbational energies, is reproduced as well as the functional form of the strong-coupling 
expansion along with accurate coefficients. Explicit accurate expressions for the ground- and first-excited state 
energies, $E_0(g)$ and $E_1(g)$ are also presented.
%%%%%%%%%%%%%%%%%%%%%%%%%%%%%%%%%%%%%%%%%%%%%%%%%%%%%%%%%%%%%%%%%%%%%%%%%%%%%%%%%%%%%%%%%%%%%%%%%%%%%%%%%%%%%%%%%%%%%%%%%%%%%%%
\end{abstract}
\noindent
\maketitle
%%%%%%%%%%%%%%%%%%%%%%
\section{Introduction}
%%%%%%%%%%%%%%%%%%%%%%
In this work we consider the familiar one-dimensional anharmonic quartic oscillator described by the Hamiltonian
\be
H = -\frac{1}{2} \dv[2]{\;}{x} + \frac{\omega^2}{2} x^2 + g x^4,
\label{H}
\ee
where $\omega^2$ and $g \ge 0$ denote the harmonic force and coupling constants, respectively.
The quartic oscillator is one of the simplest yet non-trivial system of quantum mechanics.
It is widely used as a theoretical model in many fields including 
quantum chemistry (anharmonic vibrational effects in molecular spectrocopy),
solid-sate physics, laser theory, nuclear physics, and quantum field theory.
The quartic oscillator has led to a vast literature from a mathematical, numerical, and
physical point of view. A particularly interesting feature of this elementary model is the
divergence of the Rayleigh-Schr\"odinger perturbation series of the ground-state energy for all $g>0$.
Understanding the origin of this divergence and  establishing efficient resummation techniques
for this paradigmatic divergent series is at the heart of most studies.
Among these, probably the most influential one is that of Bender and Wu\cite{Bender_1969,Bender_1971} who made 
a detailed mathematical/numerical analysis 
of the model. The authors were able to answer a number of important questions and, in particular, to shed some light
on the origin of the divergence by investigating the location of the singularities of the energy in the complex $g$-plane. Many others works have followed, as referenced below.
With respect to the divergence issue, virtually all known resummation methods have been applied to this series. The two predominant approaches 
are the use of Pad\'e approximants\cite{Baker_1975,Barry_1970,Barry_1982} and Borel's integral
summation method.\cite{Borel_1928} However, various alternative approaches have been developed, including 
the use of hypergeometric functions (\cite{Mera_2018} and references therein), nonlinear sequence of transformations,\cite{Weniger_1989,Weniger_1991} Borel transformation with a conformal
mapping,\cite{Le_Guillou_1980} or an order-dependent mapping,\cite{Seznec_1979}, as well as sequences of
analytic approximations\cite{Ivanov_1996}, among others.

In this study, we are interested in evaluating the partition function (PF) of the quartic oscillator
defined as
\be
Z = {\rm Tr} e^{-\beta H} = \sum_{n=0}^{+\infty} e^{-\beta E_n}
\label{Z}
\ee
where $E_n$ are the discretized energies of the system and $\beta$ the inverse temperature. 
A straightforward and direct method for evaluating the partition function consists in
adding up the exponential components (Boltzmann weights) of Eq.(\ref{Z}), see {\it e.g.} \cite{Schwarz_1976,Pant_1979}.
However, the exact energies being not known, approximate numerical energies are required, for example by diagonalizing the Hamiltonian
matrix within a sufficiently large basis set of Gaussian functions.\cite{Hioe_1975}
In view of the simplicity of the model, highly accurate energies can be obtained, enabling
the computation of an "exact" numerical PF, at least for not too small values of $\beta$.
In the following, we shall employ these precise numerical values as our "exact" reference.
Among the analytic methods for evaluating the PF, the oldest one is certainly the Wigner-Kirkwood perturbation
expansion of the PF in powers of $\hslash$ (or inverse temperature $\beta$), a method that systematically evaluates 
the quantum corrections to the classical partition function.\cite{Wigner_1932,Kirkwood_1933,Landau_1959}
This approach has been improved and extended in different ways, {\it e.g.} \cite{Witschel_1980},\cite{Jisba_2014}.
Another approach is the thermodynamic variation perturbation method\cite{Witschel_1983,Witschel_1992} based on a Schwinger-type expansion of the
partition function.\cite{Saenz_1955} Furthermore, exact upper and lower bounds for the PF 
have been obtained.\cite{Vinette_1991,Girardeau_1974,Falco_1995}
A natural framework to work out approximations for the PF is the path integral formalism as used in the present work. The most prominent approach following this route was pioneered by Feynman\cite{Feynman_1972} and then refined  by Feynman and
Kleinert\cite{Feynman_1986} (also, independently, by Giachetti and Tognetti\cite{Giachetti_1985}). Over the years,
the method of Feynamn and Kleinert has been systematically improved and extended
by Kleinert and collaborators\cite{Kleinert_1993,Kleinert_1994,Karrlein_1994,Kleinert_book}.

The main result of this work is to derive a closed-form expression for the partition function using a path
integral formalism. Although approximate, it captures
some of the key features of the exact partition function for all temperatures and coupling constants 
(from the weak- to the strong- coupling regimes).
From the partition function we also derive explicit expressions for the ground-
and first-excited state energies $E_0(g)$ and $E_1(g)$, respectively.
From a general point of view, our simple partition function reproducing the essential features of the exact partition function
provides an interesting model to investigate
the properties of the quartic oscillator.\\

This paper is organized as follows. In Section \ref{PF} we first derive an exact path integral expression for the PF.
Then, it is shown that at zero coupling the partition function of the harmonic oscillator is recovered.
In Section \ref{Approx} a gaussian approximation for the probability density appearing in the path integral is introduced.
In practice, this amounts to approximating the exact partition function by that 
of a harmonic oscillator with an effective frequency depending both on the temperature and coupling constant $g$.
Then, we propose in Section \ref{PMS} to invoke 
a Principle of Minimal Sensitivity (PMS) of the PF to the effective frequency.
Remarkably, this principle leads to a mathematical constraint allowing
to define a mathematically well-founded partition function, which is our final expression.
Section \ref{E0} is devoted to the derivation of explicit expressions for $E_0(g)$ and $E_1(g)$, and to the
calculation of the coefficients of the weak- and strong-coupling expansions of the ground-state energy.
In Section \ref{comparative} we present a comparative study between our formula 
and the partition functions of Feynman and Kleinert\cite{Feynman_1986} and of B\"uttner and Flytzanis\cite{Buttner_1987}.
Finally, in Section \ref{summa}, a summary of our main results and a few perspectives are presented.
%%%%%%%%%%%%%%%%%%%%%%%%%%%%%%%%%%%%%%%%%%%%%%
\section{Partition function as a path integral}
%%%%%%%%%%%%%%%%%%%%%%%%%%%%%%%%%%%%%%%%%%%%%%
\label{PF}
In the position representation the partition function writes
\be
Z = \int dx \langle x | e^{-\beta H} | x \rangle.
\ee
To obtain its path integral representation, we follow the standard route (see, {\it e.g.}, [\onlinecite{Schulman_2005})] : 
The exponential operator is broken into a product of 
$n$ exponential operators as $e^{-\beta H}=\prod_{i=1}^n  e^{-\tau H }$
with $\tau=\frac{\beta}{n}$, and 
the spectral resolution of the identity operator, $\mathds{1}=\int dx_i |x_i\rangle \langle x_i|$, is introduced between each operator giving
\be
Z = \int dx_1 ... \int dx_n  \prod_{i=1}^{n} \langle x_i | e^{-\tau H}| x_{i+1}\rangle 
\ee
where the initial and final points are identified to $x$, $x_1=x_{n+1}=x$. 
To proceed, we introduce a high-temperature (or small $\tau$) approximation of the quantity 
$\langle x_i | e^{-\tau H}| x_{i+1}\rangle$ (known as the propagator or Green's function). At the lowest order in 
$\tau$, we have
\be
\langle x_i | e^{-\tau H}| x_{i+1}\rangle = \frac{1}{\sqrt{2\pi \tau}}  e^{-\frac{(x_{i+1}-x_i)^2}{2\tau} - \tau V(x_i)}
+ O(\tau^2)
\ee
where $V(x)$ is the potential, here $V(x)= \frac{\omega^2}{2} x^2 + g x^4$. 
By taking the large-$n$ limit the contribution of the quadratic error vanishes, leading to the following exact path integral expression of 
the partition function 
\be
Z = \lim_{n \rightarrow \infty} Z_n
\ee
where the discretized partition function $Z_n$ writes
\be
Z_n=
 \qty(\frac{1}{\sqrt{2\pi \tau}})^n  \int dx_1 ... \int dx_n  
\ee
\be
e^{-\frac{1}{\tau} \sum_{i,j=1}^n x_i A_{ij} x_j - \tau \sum_{i=1}^n V(x_i)}.
\ee
Here, the matrix $A$ is given by
\be A_{ij}= \delta_{ij} - \frac{1}{2} \qty(\delta_{ii+1} + \delta_{i-1i} )
\ee
with the boundary conditions $A_{1n}=A_ {n1}=-\frac{1}{2}$.
In the following, for the sake of convenience, the $n$-dimensional integrals will be denoted as $\int d{\bf x} $, where ${\bf x} =(x_1,...,x_n)$.\\

The matrix $A$ can be diagonalized by using a Fourier transform, we get
\be
Z_n=  \qty( \frac{1}{\sqrt{2\pi \tau}})^n \int d{\bf x} e^{-\frac{1}{\tau} \sum_{i=1}^n  \lambda_i \tilde{x}^2_i}
e^{-\tau \sum_{i=1}^n V(x_i) }
\ee
where $\lambda_i$ are the eigenvalues of $A$. For $n >2$ the $\lambda_i$'s are given by
\be
\lambda_i = 1 - \cos{ \frac{2\pi}{n}(i-1)}   \;\;\; i=1 \;{\rm to}\;n.
\ee
The eigenvectors will be denoted by $\tilde{x}_i$ and decomposed as
\be
\tilde{x}_i = \sum_{j=1}^n O_{ij} x_j
\ee
where $O_{ij}$ is the orthogonal matrix diagonalizing $A$. The orthogonality condition writes
\be
\sum_{k=1}^n O_{ik} O_{jk} =\delta_{ij}.
\label{ortho}
\ee
Let us now introduce the probability density function defined as
\be
\pi(x) = \frac{e^{-\tau V(x)}}{I_v}
\label{pi}
\ee
where $I_v$ is the normalization factor
\be
I_v = \int dx e^{-\tau V(x) }.
\label{def_Iv}
\ee
The discretized partition function can be written as
\be
Z_n= \qty( \frac{I_v}{\sqrt{2\pi \tau}})^n \int d{\bf x} \prod_{i=1}^n \pi(x_i) e^{-\frac{\lambda_i}{\tau}  {\tilde x}^2_i({\bf x})}.
\label{Zex}
\ee
\\

At this point, no approximation has been made. As $n \rightarrow \infty$, $Z_n$ converges 
to the exact partition function. Unfortunately, for an arbitrary density $\pi(x)$ [equivalently, an arbitrary potential $V(x)$], the 
multi-dimensional integrals cannot be performed analytically and approximations are to be introduced. This will be the subject of the two 
following sections. 

Before doing this, let us first verify that the harmonic partition function, denoted here as $Z^{(0)}$, is recovered at zero coupling. $Z^{(0)}$ 
is given by
\be Z^{(0)}= \sum_{k=0}^{+\infty} e^{- \beta (k + \frac{1}{2}) \omega} =
	\frac{1}{ e^{ \frac{\beta \omega}{2}} - e^{-\frac{\beta \omega}{2}}}.
	\label{Zharm}
	\ee
Besides checking the validity of the expression for $Z_n$ in a particular case, the derivation of $Z^{(0)}$ to follow will also be of interest 
for the next section.\\

At $g=0$ the probability density $\pi(x)$, Eq.(\ref{pi}), is gaussian and writes
\be
\pi(x) =\frac{e^{-\tau \frac{1}{2} \omega^2 x^2 }}{\sqrt{ \frac{2\pi}{\tau \omega^2}}}
\ee
Using the orthogonality of the matrix $O$, Eq.(\ref{ortho}), leading to $\sum_{i=1}^n  {\tilde x}^2_i = \sum_{i=1}^n  x^2_i$, $Z_n$ can be written as a product of one-dimensional gaussian integrals.
Performing the gaussian ${\bf x}$-integration we get
\be
Z^{(0)}_n =  \qty( \frac{ \omega \sqrt{\tau} I^{0}_v}{\sqrt{2\pi}} )^n \qty(  \frac{1}{\omega \tau} ) \qty[ \prod_{k=2}^{n} \frac{1}{\sqrt{2 \lambda_k}}]   \prod_{i=2}^n  \frac{1}{ \sqrt{1+ \frac{\omega^2 \beta^2 }{ 2 \lambda_i n^2 }}}.
\label{Z_osc_int}
\ee
Here, $I^{(0)}_v$ is equal to $\sqrt{\frac{2\pi}{\tau \omega^2}}$, leading the first factor of the RHS of the equation to be one. 
The quantity given in brackets is equal to $\frac{1}{n}$, a result which follows from 
the relation (see, its derivation in Appendix \ref{appendix_A})
\be
\prod_{k=1}^{n-1} \sin{ \frac{k \pi }{n}} =  \frac{n}{2^{n-1}}
\ee
which leads to
\be
\prod_{k=2}^{n} \frac{1}{\sqrt{2 \lambda_k}} = \prod_{k=1}^{n-1} \frac{1}{2 \sin{ \frac{k \pi }{n}}} = \frac{1}{n}.
\ee
Now, let us  introduce the function $P_n(x)$ defined as
\be
P_n(x)= \prod_{i=2}^n  \frac{1}{\sqrt{ 1 +\frac{x^2 }{2 \lambda_i n^2}}}.
\ee
Using $P_n$ and the preceding results, the partition function can be written as
\be
Z^{(0)}_n = \frac{P_n(\beta \omega)}{\beta \omega}.
\ee
Evaluating $P_n(x)$ as $n \rightarrow \infty$ is not straightforward and requires some algrebra. In Appendix \ref{appendix_B} we show that
\be
\lim_{n \rightarrow \infty} P_n(x)=  \frac{{x}}{ e^{ \frac{{x}}{2}} - e^{-\frac{{x}}{2}}}.
\label{eqPn}
\ee
Finally,
\be
Z^{(0)}= \lim_{n \rightarrow \infty} Z^{(0)}_n = \frac{1}{ e^{ \frac{\beta \omega }{2}} - e^{-\frac{\beta \omega}{2}}},
\ee
in agreement with Eq.(\ref{Zharm}).
Note that the derivation of the partition function of the harmonic oscillator and, more generally, of the 
Green's function  $K(x,x^{\prime},\beta)= \langle x| e^{-\beta H}| 
x^{\prime} \rangle$, has been done in the literature in many different ways, see for example \cite{Feynman_1965,Schulman_2005,Marshall_1979,Cohen_1998,Holstein_1998,Barone_2003,Hira_2013}. The present derivation is related to that of Cohen\cite{Cohen_1998}.
%%%%%%%%%%%%%%%%%%%%%%%%%%%%%%%%
\section{Gaussian approximation}
%%%%%%%%%%%%%%%%%%%%%%%%%%%%%%%%
\label{Approx}
The first approximation introduced here for evaluating $Z_n$, Eq.(\ref{Zex}), 
consists in approximating the general density $\pi(x)= \frac{e^{-\tau V(x)}}{\int dx e^{-\tau V(x)}}$ 
by a gaussian density, $\pi_G$, corresponding to an effective harmonic oscillator of frequency $\omega_g(\tau)$
\be
\pi(x) \sim \pi_G(x)=\frac{ e^{-\tau \frac{1}{2} \omega^2_g(\tau) x^2 }}{\int dx e^{-\tau \frac{1}{2} \omega^2_g(\tau) x^2}}
\label{gauss_approximation}
\ee
Note that when the coupling constant $g$ goes to zero, the effective frequency 
reduces to $\omega$.
Different criteria can be chosen to define the optimal effective frequency minimizing the 
error in the approximation. Here, we propose to impose the variance of $\pi_G$ to be equal to the exact one, that is
\be
\sigma^2(\tau)= \langle x^2 \rangle_{\pi_G} = \langle x^2 \rangle_{\pi}.
\label{sigma2}
\ee
After some manipulations, this equality leads to
\be
\omega_g(\tau) = \omega \sqrt{B(\frac{4 g}{\tau \omega^4})}
\label{eqbeta}
\ee
where the parameter-free function $B(x)$ is defined as
\be
B(x) = \frac{1}{2} \frac{\int dy \; e^{-y^2 - x y^4}} { \int dy \; y^2 e^{-y^2 - x y^4}},
\label{beta}
\ee
a function which can be expressed as
\be
B(x) = \frac{ 4x K_{\frac{1}{4}}\qty(\frac{1}{8x})}
{K_{-\frac{3}{4}}\qty(\frac{1}{8x}) + K_{\frac{5}{4}}\qty(\frac{1}{8x}) - 2 (1+2x) K_{\frac{1}{4}}(\frac{1}{8x})}
\ee
where $K_\nu(x)$ is the modified Bessel function of the second kind.
The function $B(x)$ is positive and monotonically increasing. It starts at $x=0$ with $B(0)=1$, then 
increases linearly at small $x$'s and, finally, behaves as $\sim \sqrt{x}$ at large $x$.\\

The gaussian approximation being made, $Z_n$ becomes the partition function of a harmonic oscillator
\be
Z_n= \qty(\frac{I_v}{\sqrt{2\pi\tau}} )^n  
	\int d{\bf x} \prod_{j=1}^n \frac{ e^{-\frac{\tau \omega^2_g(\tau) x^2_j}{2}} } 
	{ \int dx_j  e^{-\frac{\tau \omega^2_g(\tau) x^2_j}{2}} } 
\label{gaussian_form}
\ee
which can be evaluated as done in the previous section for the usual harmonic oscillator.
After the ${\bf x}$ gaussian integration,  
$Z_n$ takes the form of Eq.(\ref{Z_osc_int}), with the replacements
\be
I^{0}_v \rightarrow I_v  \;\;\; {\rm and} \;\;\; \omega \rightarrow  \omega_g(\tau).
\ee

The partition function is then given by
\be
Z_n= \qty(\frac{\omega_g(\tau) \sqrt{\tau} I_v}{ \sqrt{2\pi} } )^n \frac{1}{\beta \omega_g(\tau)} 
P_n\qty[\beta \omega_g(\tau)].
\label{Zapprox0}
\ee
Finally, by performing the limit $n \rightarrow \infty$ for $P_n\qty[\beta \omega_g(\tau)]$ only, we are led to the following expression for $Z_n$ 
\be
Z_n= \qty(  \frac{\omega_g(\tau) \sqrt{\tau} I_v}{ \sqrt{2\pi} } )^n  \frac{1}{ e^{\frac{{\beta \omega_g(\tau)}}{2}}-e^{-\frac{{\beta \omega_g(\tau)}}{2}}}.
\label{Zapprox}
\ee
Note that, since $P_n$ has been replaced by $P_{\infty}$, this formula is now only valid in the large-$n$ limit.
In the harmonic case, the quantity $C(\tau) \equiv \frac{\omega_g(\tau) \sqrt{\tau} I_v}{ \sqrt{2\pi} }$  
is equal to one and $\omega_g(\tau)=\omega$. $Z_n$ thus becomes independent of $n$
and the infinite-$n$ limit becomes trivial.
At non-zero coupling, this is no longer the case. In the large-$n$ limit, we have 
$\omega_g(\tau) \sim {\tau}^{-\frac{1}{4}}$ and $I_v \sim {\tau}^{-\frac{1}{4}}$. As a consequence, in the small-$\tau$ 
limit $C(\tau)$ 
converges to a finite constant independent of $\omega$ and $g$, given by 
$C(0)=2\sqrt{\frac{2}{\pi}} \frac{\Gamma^{\frac{3}{2}}(\frac{5}{4}) }{\Gamma^{\frac{1}{2}}(\frac{3}{4})}$ = 1.24397...
This constant being greater than 1, the partition function diverges geometrically as $Z_n \sim C(0)^n e^{-\alpha\; n^{\frac{1}{4}}}$ 
with $\alpha > 0$
\be
Z_n \rightarrow  +\infty  \;\; {\rm as} \;\; n \rightarrow +\infty.
\label{noconvergence}
\ee
In the next section we propose to overcome this problem by introducing an additional constraint to $Z_n$.
%%%%%%%%%%%%%%%%%%%%%%%%%%%%%%%%%%%%%%%%%%%%%%%%%%%%%%%%%%
\section{Principle of Minimal sensitivity}
%%%%%%%%%%%%%%%%%%%%%%%%%%%%%%%%%%%%%%%%%%%%%%%%%%%%%%%%%%
\label{PMS}
In the preceding section, the exact partition function has been approximated by the partition function of a harmonic oscillator 
with an effective frequency $\omega_g(\tau)$.
Now, the exact partition function being independent of $\omega_g(\tau)$ it is natural to impose a minimal sensitivity of $Z_n$ 
to the choice of this frequency. Mathematically, this condition writes
\be
\frac{ \partial Z_n}{\partial \omega_g(\tau)} = 0.
\label{cond}
\ee
Using the expression of $Z_n$ given by Eq.(\ref{Zapprox}),
the stationarity condition leads to the following implicit equation for the variable $n$
\be
n=\frac{ \beta \omega_g\qty(\frac{\beta}{n}) }{2} {\coth{\frac{\beta \omega_g\qty(\frac{\beta}{n}) }{2}}}.
\label{eq0}
\ee
As shown below, this equation actually admits a {\it unique} solution denoted, here, as $n=n_c(\beta)$ (in general, not an integer).
We thus see that imposing to the gaussian approximation of the exact $Z_n$ both a minimal error 
in the approximation and a 
minimal sensitivity to the choice of the effective frequency turns out to be possible only for a single value of $n=n_c(\beta)$.
We thus propose to define $Z$ as
\be
Z=Z_{n_c(\beta)}.
\ee
The unicity of the solution is shown as follows. Let us rewrite Eq.(\ref{eq0}) as $f(n)=g(n)$ where $f(n)=n$ 
and $g(n)=\frac{ \beta \omega_g\qty(\frac{\beta}{n}) }{2} {\coth{\frac{ \beta \omega_g\qty(\frac{\beta}{n}) }{2}}}$.
Both functions are monotonically increasing functions of $n$. In the case of the function $g(n)$ this is true because 
$\omega_g(\tau)$ is an increasing function of $n$.
At large $n$, $f(n)$ increases linearly, while $g(n)$ increases 
slower as $\sim n^{\frac{1}{4}}$.
Now, the function $f$ starting with a smaller value than $g$, $f(0)=0 < g(0)$, and increasing more rapidly,
the two functions necessarily cross at a unique value of $n$.\\

To summarize, our final formula for the partition function of the quartic oscillator is given as follows.
Introducing the convenient $\beta$-dependent effective $\tau$ denoted as $\tau_c(\beta)$ and defined as
\be
\tau_c(\beta)= \frac{\beta}{n_c(\beta)},
\ee
the partition function writes
\be
Z= \qty(  \frac{ \omega_g\qty[\tau_c(\beta)] \sqrt{\tau_c(\beta)} I_v\qty[\tau_c(\beta)] }{\sqrt{2\pi}}  )^{n_c(\beta)}
\frac{1}{ e^{\frac{\beta \omega_g\qty[\tau_c(\beta)] }{2}}
-e^{-\frac{\beta \omega_g\qty[\tau_c(\beta)] }{2}}}
\label{Zdef}
\ee
with
\be
I_v\qty[\tau_c(\beta)]= \int dx e^{-\tau_c(\beta) V(x)}.
\label{Zdef_I}
\ee
The quantities $n_c(\beta)$ and $\tau_c(\beta)=\frac{\beta}{n_c(\beta)}$ are obtained by solving (for example, iteratively) the 
pair of equations
\be
n_c(\beta)=\frac{ \beta \omega_g\qty[\tau_c(\beta)] }{2} {\coth{\frac{\beta \omega_g\qty[\tau_c(\beta)] }{2}}}
\label{eq1}
\ee
and
\be
\omega_g\qty[\tau_c(\beta)]= \omega \sqrt{B\qty[\frac{4g}{\tau_c(\beta) \omega^4}]}.
\label{eq2}
\ee
Note that, besides reducing to the partition function of the harmonic oscillator at zero coupling, our expression also 
reduces to the classical partition function
in the high-temperature limit, $\beta \rightarrow 0$. Indeed, in this limit
we have $n_c(\beta) \rightarrow 1$, $\tau_c(\beta) \rightarrow \; \sim \beta$, and $\omega_g(\beta) \rightarrow \; \sim \beta^{-\frac{1}{4}}$
leading to the exact classical partition function
\be
Z \rightarrow  \frac{1}{\sqrt{2\pi \beta}} \int dx e^{-\beta V(x)}.
\label{Z_clas}
\ee

\begin{figure}[htb]
\centering
\includegraphics[scale=0.40]{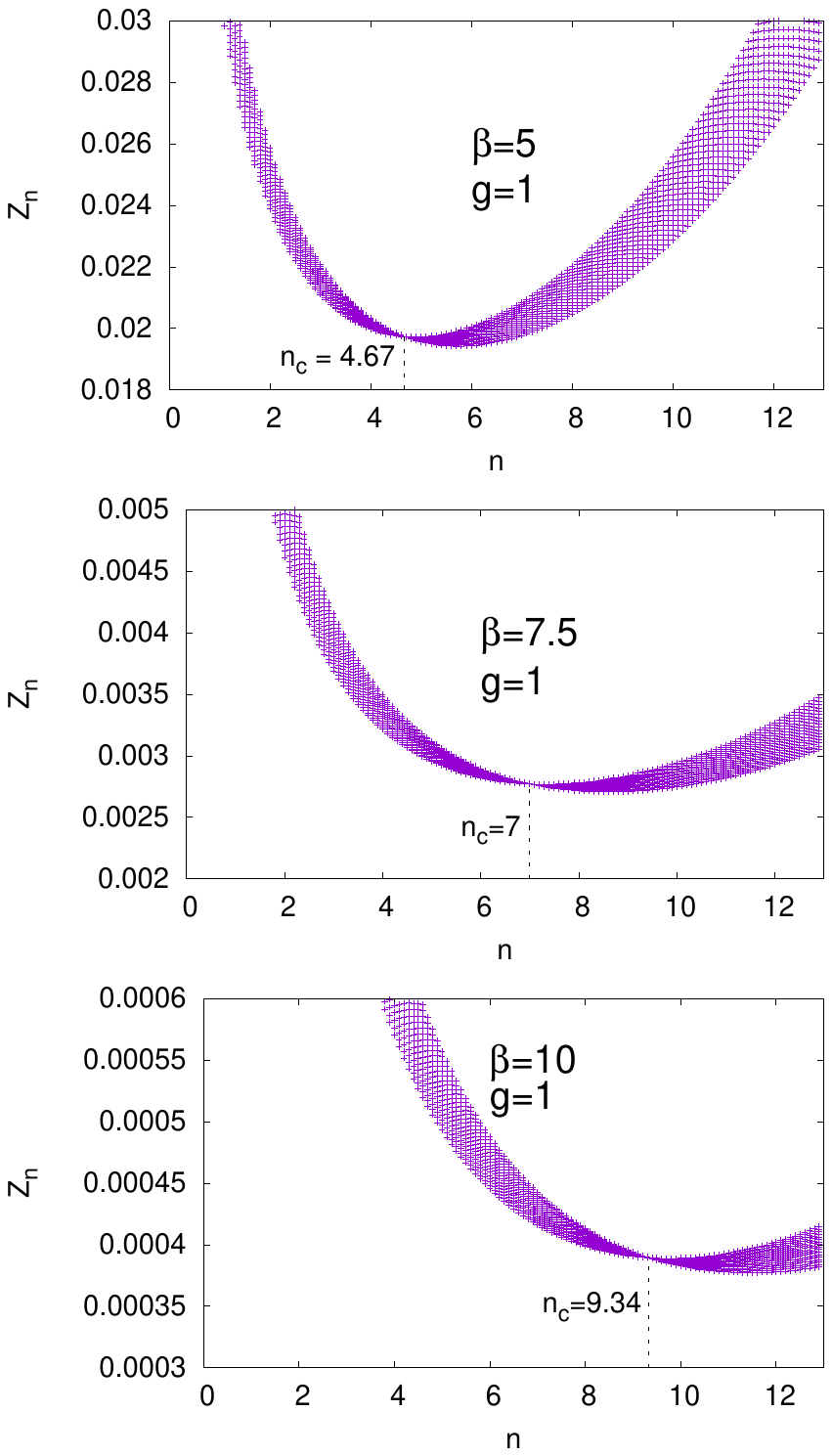}
\caption{$Z_n$, Eq.(\ref{Zapprox}), as a function of $n$ for $\omega=1$, $g=1$, and $\beta=5, 7.5$, and $10$. 
For each $n$, $Z_n$ is evaluated for thirteen different values of $\omega_g$'s uniformly distributed
around $\omega_g(\tau)$ as given by Eq.(\ref{eqbeta}).}
\label{fig1}
\end{figure}

\begin{figure}[htb]
\centering
\includegraphics[scale=0.40]{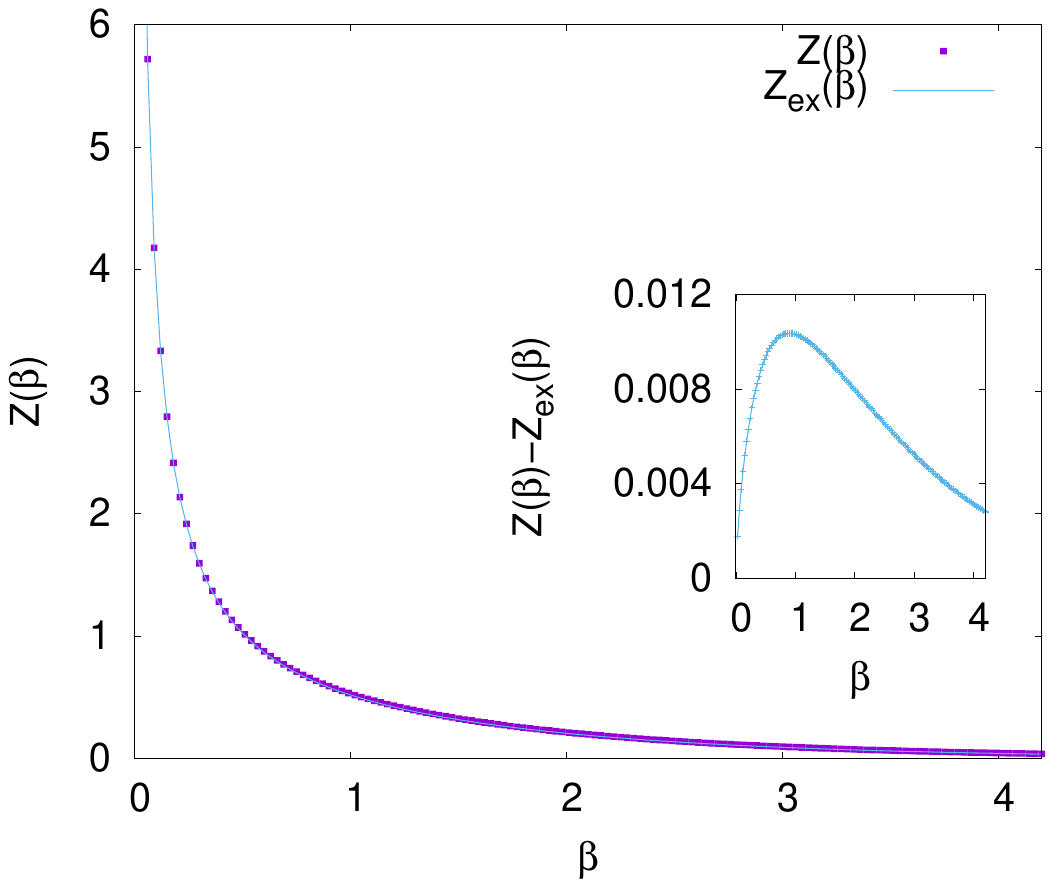}
	\caption{Comparaison between $Z(\beta)$ and the exact 
	numerical partition function $Z_{\rm ex}(\beta)$. The inset shows the error. Here, $\omega=1$ and $g=1$.}
\label{fig2}
\end{figure}

Introducing the principle of minimal sensitivity being a critical step of this work, 
it is worth presenting some numerical calculations illustrating this principle.
In figure \ref{fig1} the partition function $Z_n$ given by Eq.(\ref{Zapprox}), that is, 
before the principle of minimal sensitivity has been introduced,
is shown as a function of $n$ for three different values of the inverse temperature,
$\beta=5, 7.5$, and $10$ and for $\omega=g=1$. For each $\beta$ and $n$,
$Z_n$ is evaluated for thirteen different values of the effective frequency $\omega_g$.
For each $n$, the $\omega_g$'s have been chosen to be uniformly distributed 
around $\omega_g(\tau)$, the optimal effective frequency leading to a minimal error in the gaussian approximation, 
as given by Eq.(\ref{eqbeta}).
Figure \ref{fig1} illustrates the fact that $Z_n$ diverges 
at large $n$, Eq.(\ref{noconvergence}). We also see that, for each $n$, $Z_n$ 
is strongly sensitive to the value of $\omega_g$ used, except at a unique value of 
$n$ [=$n_c(\beta)$] where the stationarity condition, $\frac{\partial Z_n}{\partial \omega_g(\tau)} = 0$ holds.
The values of $n_c$ observed on the figure
coincide with those obtained by solving the pair of equations, Eqs.(\ref{eq1}) and (\ref{eq2}), as it should.
These various results, discussed here only for $g=1$ and three different $\beta$'s, have been found 
to be valid for any value of $\beta$ and $g$ (data not shown here).
%%%%%%% FIG1 %%%%%%%%
%%%%%%%%%%%%%%%%%%%%
In figure \ref{fig2} we compare the partition function $Z$
as a function of $\beta$ and the exact one obtained by explicit 
summation of the exponential components of $Z$ using highly accurate numerical energies (solid line in the figure). 
As seen, at the scale of the figure, both sets of data are nearly indistinguishable. To increase the resolution, the inset presents
the difference $Z(\beta)-Z_{\rm ex}(\beta)$. The maximum error is about 0.01. 
As expected, the error vanishes in the high-temperature limit, $\beta \rightarrow 0$.
%%%% FIG2 %%%%%%%%%
%%%% FIG2 %%%%%%%%%

%%%%%%%%%%%%%%%%
\section{Energy}
%%%%%%%%%%%%%%%%
\label{E0}
In this section the formulas for the ground- and first-excited energies, $E_0(g)$ and $E_1(g)$ are derived.
Next, the coefficients of the weak- and strong-coupling expansions of the ground-state energy are evaluated and compared to the 
exact ones. 
%%%%%%%%%%%%%%%%%%%%%%%%%%%%%%%%%%%%%%%%%%%%%%
\subsection{Ground- and first-excited energies}
%%%%%%%%%%%%%%%%%%%%%%%%%%%%%%%%%%%%%%%%%%%%%%
The ground-state energy is obtained from the zero-temperature limit of the free energy as follows
\be
E_0(g) = \lim_{\beta \rightarrow \infty} -\frac{1}{\beta} \ln{Z}.
\label{e0g}
\ee
At large $\beta$, the (unique) solution of Eqs.(\ref{eq1}) and (\ref{eq2}) is found to be proportional to $\beta$
\be
n_c(\beta)=\frac{{\bar \omega_g}}{2} \beta,
\ee
where  ${\bar \omega_g}$ denotes the solution of the implicit equation given by
\be
{\bar \omega_g}= \omega \sqrt{B\qty( \frac{2g {\bar \omega_g}}{\omega^4})}.
\label{implicit}
\ee
Note that the proportionality of $n_c(\beta)$ with $\beta$ is consistent with the data presented in Fig.\ref{fig1}.
Using the zero-temperature limit of the free energy and the expression 
for the partition function, we get
\be
E_0(g) = \frac{ {\bar \omega_g} }{2} \qty[ 1 -\ln{ \qty(\sqrt{\frac{\bar \omega_g}{\pi}} {\bar I}(g))} ]
\label{E0_final}
\ee
where
\be
{\bar I}(g)= \int dx e^{ -\frac{2}{{\bar \omega_g}} V(x)}.
\label{Ig}
\ee
To get the first-excited energy $E_1(g)$, the 
first subleading contribution to $Z(\beta)$ at large $\beta$ has to be evaluated. For that, we first 
rewrite the equation obeyed by ${\omega_g}(\beta)$ under the form
\be
\omega_g(\beta) = \omega \sqrt{
B\qty[ \frac{2g}{\omega^4}  {\omega_g}(\beta) \coth{ \frac{\beta {\omega_g}(\beta)}{2} }
\label{eqomeg}
]}.
\ee
Introducing the variable $y$ defined as
\be
y=e^{-\beta \omega_g}
\ee
the hyperbolic cotangent can be expanded for all values of $y <1$ as
\be
\coth{\frac{\beta {\omega_g}(\beta)}{2}} = 1 + 2 \sum_{n=1}^{\infty} y^n.
\ee
As a consequence of Eq.(\ref{eqomeg}) and of the fact that the function $B(x)$ is infinitely differentiable, 
the effective frequency can also be expanded in powers of $y$. In addition, 
at the lowest order $y \sim e^{-\beta {\bar \omega_g}}$ up to an exponentially small 
correction. By performing the Taylor expansion of $B(x)$ at the value  $\frac{2g}{\omega^4} {\bar \omega_g}$ we get at the lowest order
\be
\omega_g(\beta) = {\bar \omega_g} + {\bar c}_g  e^{-\beta {\bar \omega_g}} + O\qty[ e^{-2\beta {\bar \omega_g}}]
\ee
with
\be
{\bar c}_g= \frac{2 u_g}{1 - \frac{u_g}{ {\bar \omega}_g }}
\ee
and
\be
u_g = \frac{g}{\omega^2} B^\prime \qty(\frac{2g}{\omega^4} {\bar \omega}_g).
\ee
Incorporating the first-order expression for $\omega_g(\beta)$ into the expression of the partition function, 
it is also possible to expand the PF with respect to $y$. 
After some algebra the two first leading contributions to the PF write
\be
Z(\beta) = e^{-\beta E_0} + e^{-\beta \qty(E_0 + {\bar \omega}_g)} \qty[1- \beta \Delta\qty(g)] + O\qty[ e^{-\beta \qty(E_0 +2 {\bar \omega}_g)} ]
\ee
with
\be
\Delta(g)= \qty(1 +\frac{{\bar c_g}}{2 {\bar \omega_g}} ) \qty[ 1 - \frac{ 4{\bar V(g)}}{ {\bar \omega_g} {{\bar I}(g)}}
- 2 \ln{ \qty( \sqrt{\frac{{\bar \omega_g}}{\pi}} {{\bar I}(g)} )} ]
\ee
and 
\be
{\bar V}(g) = \int dx V(x) e^{ -\frac{2}{{\bar \omega_g}} V(x)}.
\ee
Due to the presence of the linear term $\qty[1-\beta \Delta\qty(g)]$ in $Z(\beta)$, 
our approximate partition function cannot be expressed as an infinite sum of Boltzman factors as it 
should be for the exact one. As a consequence, in order to define what is meant by excited-state energies
in our model, we need to introduce some prescription. Here, we propose to exponentiate the linear contribution 
$\qty[1-\beta \Delta\qty(g)]$ to give $e^{-\beta \Delta(g)}$. 
By doing this, the first subleading contribution to $Z(\beta)$ becomes a pure exponential 
and the first excited-state energy is defined as
\be
E_1(g) = E_0(g)+{\bar \omega_g} + \Delta(g).
\ee
Note that, in practice, the exponentiation has a marginal quantitative impact since
the product $\beta \Delta(g)$ remains small for typical values of $\beta$.
Preliminary investigations show that a generalization of such a strategy for evaluating higher excited-state energies
is possible. However, the general structure is not so simple. This study is left 
for future research.

\begin{figure}[htb]
\centering
	\includegraphics[scale=0.40]{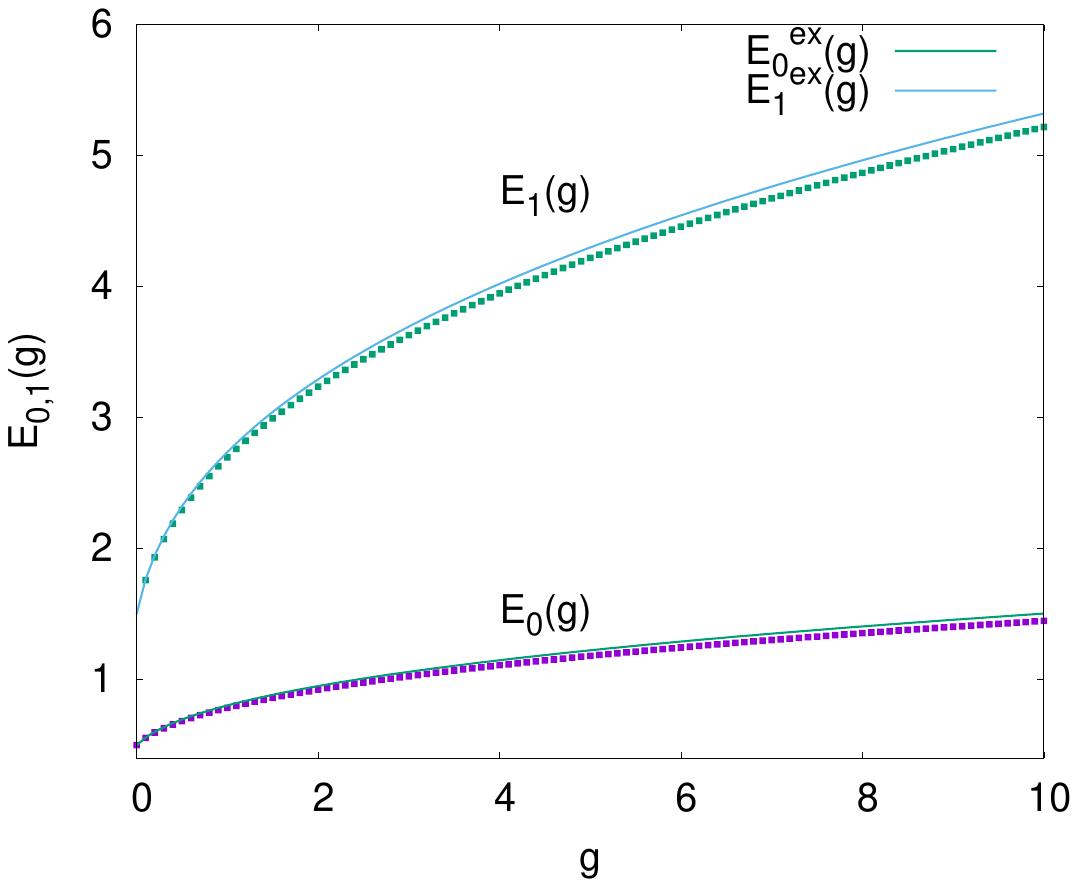}
\caption{$E_0$ and $E_1$ as a function of $g$. Exact results given by the solid lines. $\omega=1$.}
\label{fig3}
\end{figure}

In figure \ref{fig3} the variation of $E_0$ and $E_1$ as a function of $g$ for $\omega=1$ is shown.
The exact curves are represented by solid lines. At the scale of the figure, $E_0(g)$ is in excellent agreement with the exact energy 
over the full range of $g$. The first excited-energy is 
slighlty less accurate and has also the correct overall behavior. Note that both $E_0$ and $E_1$ are systematically smaller than the exact ones for 
all values of the coupling constant. 
%%%%% FIG3 %%%%%%%%
%%%%%%%%%%%%%%%%%%
%%%%%%%%%%%%%%%%%%%%%%%%%%%%%%%%%%%%%%%
\subsection{Weak coupling expansion}
%%%%%%%%%%%%%%%%%%%%%%%%%%%%%%%%%%%%%%%
For small values of $g$ (weak-coupling) the Rayleigh-Schr\"odinger
perturbation theory expresses the ground-state energy as a power series
\be
E_0(g)= \sum_{n=0}^{+\infty} E^{(n)} g^n.
\label{RSPT}
\ee
The $n$-th order perturbational energies $E^{(n)}$ are known to be rational numbers. For $\omega=1$ they are given by 
$\frac{1}{2},\frac{3}{4},-\frac{21}{8},\frac{333}{16},-\frac{30885}{128}, ...$. In the large-$n$ regime they
have the following asymptotic behavior\cite{Bender_1969}
\be
E^{(n)} \sim_{n \rightarrow \infty}
-\frac{\sqrt{6}}{\pi} {(\frac{-3}{4})}^n \frac{(2n)!}{n!}.
\label{An}
\ee
Because of the factorial increase of the coefficients the series has a zero radius of convergence.
In practice, by summing a finite number of components of this divergent series the ground-state energy 
can be accurately obtained only for (very) small values of $g$.
For larger values, it is necessary to use one of the methods to sum up divergent series, see references in the introduction.\\

Let us now evaluate the weak-coupling expansion of our approximate ground-state energy $E_0(g)$, Eqs.(\ref{E0_final}) and (\ref{Ig}).
At zero coupling, ${\bar \omega_g} = \omega$, and the zero-th order of the expansion reduces 
to the ground-state energy of the harmonic oscillator,
\be
E^{(0)} = \frac{\omega}{2}.
\ee
The ground-state energy depending only on ${\bar \omega_g}$, let us expand it in powers of $g$
\be
{\bar \omega_g} = \sum_{n=0}^{+\infty} u_n g^n \;\;\; {\rm with} \;\;\; u_0= \omega.
\ee
Using the implicit equation determining ${\bar \omega_g}$, Eq.(\ref{implicit}), 
the coefficients $u_n$ can be expressed in terms of the coefficients $B_n$ of the polynomial expansion of $B(x)$, 
Eq. (\ref{beta}).
Now, let us define the function $J(g)$ as
\be
J(g)=\sqrt{ \frac{{\bar \omega_g}}{\pi}} {\bar I}(g) = \frac{1}{\sqrt{\pi}} \frac{{\bar \omega_g}}{\omega} 
\int dx e^{-x^2} e^{-\frac{ 2 g {\bar \omega_g}}{\omega^4} x^4}
\ee
and expand it in power series of $g$
\be
J(g) = \sum_{n=0}^{+\infty} J_n g^n   \;\;\; {\rm with} \;\;\; J_0=1.
\ee
The coefficients are given by
\be
J_n = \frac{1}{\sqrt{\pi}} \frac{ {\bar \omega_g}^{n+1}}{\omega^{4n+1}} \frac{ \qty(-2)^n  [4n]}{n!}
\ee
where the symbol $[4n]$ is defined as
\be
[4n] =  \int dx e^{-x^2}  x^{4n} = \frac{ (4n-1)!!}{2^{2n}} \sqrt{\pi}.
\ee
Using the expansion of ${\bar \omega_g}$ the coefficients $J_n$ can be expressed in terms of the coefficients ${\bf u}=\qty(u_1, u_2, ...)$. We 
then have for the energy
\be
E_0(g) = \frac{1}{2} \qty( \sum_{n=0}^{+\infty} u_n g^n) \qty[  1 - \ln{ \qty( 1 + \sum_{n=1}^{+\infty} J_n\qty[{\bf u}] g^n )}].
\ee
As seen, the power expansion of the energy can be expressed only in terms of the coefficients ${\bf u}$. The evaluation of these coefficients
is rather tedious but can be easily performed using a symbolic computation program such as Mathematica\cite{Mathematica}.
In Table \ref{tab1} we report the exact and approximate coefficients of the power series up to $n=5$. 
Quite interestingly, the approximate coefficients are found to be rational numbers like in the exact case, the 
two first coefficients being identical. Finally, we note that
the two series display a similar mathematical pattern.
Indeed, the magnitude of the coefficients increases dramatically with the perturbational order in both 
cases (factorially, for the exact series) 
with an even more rapid rate for the approximate
series and, furthermore, the sign pattern of the coefficients is identical. 
%%%%%%% TAB1 %%%%%%%%%%%%%%%%%%%%%%%%%%%%%%%%%%%%%%%%%%%%%%%%%%%%%%%%%%%%%%%%%%%%%%%%%%
\begin{table}[htb]
\centering
\caption{Coefficients of the power series of $E_0(g)$ up to $n=5$. The first
column reports the exact coefficients. The second column the coefficients obtained in the present work.}
	\label{tab1}
\begin{tabular}{lcc}
\hline
	$	n$ &  $E^{(n)}_{{\rm ex}}$      &  $E^{(n)}$ \\
\hline
	0 &   1/2=    0.5      &      1/2=      0.5  \\
	1 &  3/4=     0.75     &      3/4=      0.75 \\
	2 &  -21/8=    -2.625    &    -15/4=       -3.75 \\
	3 &  333/16=    20.8125   &           54 \\
	4 & -30885/128=   -241.289... &  -20817/16=      -1301.0625 \\
	5 & 916731/256=   3580.98...  & 216243/5=       43248.6 \\
\hline
\end{tabular}
\end{table}
%%% END TAB 1 %%%%%%%%%%%%%%%%%%%%%%%%%%%%%%%%%%%%%%%%%%%%%%%%%%%%%%%%%%%%%%%%%%%%%%%%%%%%%%
\subsection{Strong coupling expansion}
%%%%%%%%%%%%%%%%%%%%%%%%%%%%%%%%%%%%%%%%%%%%%%%%%%%%%%%%%%%%%%%%%%%%%%%%%%%%%%%%%
In the strong-coupling regime (large $g$) it has been shown that the expansion in terms of $\frac{1}{g}$
takes the form\cite{Janke_1995_B}
\be
E_0(g)= g^{\frac{1}{3} }\sum_{n=0}^{+\infty} \alpha_n \qty(\frac{1}{g})^{\frac{2n}{3}}.
\label{strong}
\ee
Accurate values for the coefficients $\alpha_n$ have been reported.\cite{Janke_1995_B}\\

Here, to build the strong-coupling expansion of $E_0(g)$ we first need to evaluate the large-$x$ expansion of $B(x)$. From the 
definition of $B(x)$, Eq.(\ref{beta}), we easily show that (see, Appendix \ref{appendix_C})
\be
B(x)= \sqrt{x} \sum_{n=0}^{+\infty} B_n \qty(\frac{1}{\sqrt{x}})^n.
\ee
The three first values of $B_n$ are given in Appendix \ref{appendix_C}.
Now, by using the implicit equation for ${\bar \omega_g}$, 
Eq.(\ref{implicit}), we find by simple inspection that 
\be
{\bar \omega_g}(g)= g^{\frac{1}{3} }\sum_{n=0}^{+\infty} \omega_n \qty(\frac{1}{g})^{\frac{2n}{3}}.
\ee
The leading coefficient $\omega_0$ is given by
\be
\omega_0= \qty( \sqrt{2} B_0)^{\frac{2}{3} } g^{\frac{1}{3}}
\ee
with
\be
B_0= 2 \frac{ \Gamma \qty(\frac{5}{4})} {\Gamma \qty(\frac{3}{4})}.
\ee
The coefficients $\omega_1$ and $\omega_2$ are reported in Appendix \ref{appendix_C}.
Now, the strong coupling expansion of the energy may be obtained from 
the expansion of ${\bar \omega_g}(g)$ and the expression of $E_0(g)$, Eq.(\ref{E0_final}). 
We get
\be
E_0(g)= g^{\frac{1}{3} }\sum_{n=0}^{+\infty} \alpha_n \qty(\frac{1}{g})^{\frac{2n}{3}}.
\ee
Remarkably, the functional form of the exact energy, Eq.(\ref{strong}), is recovered by our model.
The explicit expression of the first three coefficients $\alpha_n$ are given in Appendix \ref{appendix_C}. Their numerical values are reported in 
Table \ref{tab2} 
and compared to the exact ones calculated in \cite{Janke_1995_B}. The relative error on the leading coefficient $\alpha_0$ is only 
about 4\%, a remarkable result in view of the simplicity of the model. The accuracy of the next coefficients decreases as a function of $n$ 
but the values are still reasonable.
%%%% TAB2 %%%%%%%%%%%%%%%%%%%%%%%%%%%%%%%%%%%%%%%%%%%%%%%%%%%%%%%%%%%%%%%%%%%%%%%%%%%%%%%%%%%%%%%%%%%%%%%%%%%%%%%%%%
\begin{table}[htb]
\centering
\caption{Strong-coupling coefficients $ \alpha_n$ compared to the exact values of \cite{Janke_1995}. Here, $\omega=1$.}
\label{tab2}
\begin{tabular}{lcc}
\hline
$n$ &  $\alpha_n$       & $\alpha^{ex}_n$ \\
0 &       0.6393...     & 0.6679...  \\
1 &       0.1576...     & 0.1436...  \\
2 &      -0.0152...     &-0.0086...  \\
\hline
\end{tabular}
\end{table}
%%%% END TAB2 %%%%%%%%%%%%%%%%%%%%%%%%%%%%%%%%%%%%%%%%%%%%%%%%%%%%%%%%%%%%%%%%%%%%%%%%%%%%%%%%%%%%%%%%%%%%%%%%%%%%%%%%%%
%%%%%%%%%%%%%%%%%%%%%%%%%%%%%%%%%%%%%%%%%%%%%%%%%%%%%%%%%%%%%%%%%%%%%%%%%%%%%%%%%%%%%%%%%%%%%%%%%%%%%%%%
\section{Comparison with the Feynman-Kleinert and B\"uttner-Flytzanis approximate partition functions}
%%%%%%%%%%%%%%%%%%%%%%%%%%%%%%%%%%%%%%%%%%%%%%%%%%%%%%%%%%%%%%%%%%%%%%%%%%%%%%%%%%%%%%%%%%%%%%%%%%%%%%%%
\label{comparative}
Two simple yet accurate partition functions for the anharmonic oscillator proposed in the literature are those of 
Feynman and Kleinert\cite{Feynman_1986} and of B\"uttner and Flytzanis\cite{Buttner_1987}. 
In this section, the respective quality of the three approximate formulas is evaluated.
%%%%%%%%%%%%%%%%%%%%%%%%%%%%%%%%%%%%%%%%%%%%%%
\subsubsection{Feynman-Kleinert (FK) approach}
%%%%%%%%%%%%%%%%%%%%%%%%%%%%%%%%%%%%%%%%%%%%%%
In the same way as in the present work, the authors make use of the path integral formalism.
However, the route followed is very different. 
In short (for details, see \cite{Feynman_1986}) the starting idea is to 
rewrite $Z$ as a classical partition function involving
an effective classical potential $V_{\rm eff}(x)$,
\be
Z= \frac{1}{\sqrt{2\pi \beta}} \int dx_0 e^{-\beta V_{\rm eff}(x_0)}.
\ee
This representation is exact and particularly well-suited to the high-temperature limit where
the formula reduces to the classical partition function, Eq.(\ref{Z_clas}).
In this limit the only path contributing to the partition function is the constant path connecting the initial and final points, that is, 
$x(0)=x(\beta)=x_0$. At small $\beta$, it is useful to introduce the collective (centro\"{\i}d) variable, 
$\bar{x} = \int_0^\beta ds x(s)$, since the paths contributing the most 
to the path integral are those 
for which $x(s)$ does not fluctuate too much around $\bar{x}$. The deviations from $\bar{x}$ are then locally approximated in a harmonic 
way, thus generating an approximation for the effective classical potential that can be explicitly obtained
by solving a pair of coupled equations. In 
Appendix \ref{appendix_D} the working formulae giving the effective potential are reported. 
From the partition function, the approximate ground-state energy can be obtained (see the derivation in 
Appendix \ref{appendix_D}). We have
 \be
 E_0(g)= \frac{\Omega_0(g)}{4} + \frac{\omega^2}{4\Omega_0(g)} + \frac{3g}{4\Omega^2_0(g)}
 \ee
 with
 \be
 \Omega_0(g)= \frac{\omega^2}{ 3^{\frac{1}{3} } \Delta(g)^{\frac{1}{3}}} + \frac{ \Delta(g)^{\frac{1}{3}}} { 3^{\frac{2}{3}}}
 \ee
and
 \be
 \Delta(g)= 27 g + \sqrt{3} \sqrt{ 243 g^2 - \omega^6}.
 \ee
 To obtain the first excited-state energy, $E_1(g)$, reported in the figure \ref{fig4} below,
the subleading component of the FK partition function, $Z_{FK}$, is extracted as follows
 \be
 E_1(g) = \lim_{\beta \rightarrow \infty} -\frac{1}{\beta} \log\qty(Z_{FK} - e^{-\beta E_0(g)}).
 \ee
 %%%%%%%%%%%%%%%%%%%%%%%%%%%%%%%%%%%%%%%%%%%%%%%%
 \subsubsection{B\"uttner-Flytzanis (BF) approach}
 %%%%%%%%%%%%%%%%%%%%%%%%%%%%%%%%%%%%%%%%%%%%%%%%%%
The B\"uttner-Flytzanis approach is based on a variational approach for the free energy, $F=-\frac{1}{\beta} \ln{Z}$. 
As noted by Feynman,\cite{Feynman_1972} an approximation 
of the free energy is the following upper limit for $F$
\be
 \tilde{F}= F_0 + \langle H-H_0 \rangle_0
\ee
where $F_0$ is the free energy of some reference Hamiltonian $H_0$ and the average is taken with respect to $H_0$. 
Using as reference Hamiltonian a harmonic oscillator of frequency ${\tilde \omega}$ with its center displaced by a quantity $b$ we have
\be
\tilde{F}= F_0 + \langle V(x) - \frac{1}{2} {\tilde \omega}^2 (x-b)^2  \rangle_0.
\ee
$\tilde{F}$ can be evaluated analytically and its minimization with respect to the 
variational parameters ${\tilde \omega}$ and $b$ leads to $b=0$ and to a third-order polynomial equation giving the optimal 
value for
${\tilde \omega}$. For completeness, the equations are presented in the Appendix \ref{appendix_E}.
%%%%%%%%%%%%%%%%%%%%%%%%%%%%%%%%%%%%%%%%%%
\subsubsection{Compararison between models}
%%%%%%%%%%%%%%%%%%%%%%%%%%%%%%%%%%%%%%%%%%
%%%% TAB3 %%%%%%%%%%%%%%%%%%%%%%%%%%%%%%%%%%%%%%%%%%%%%%%%%%%%%%%%%%%%%%%%%%%%%%%%%%%%%%%%%%%%%%%%%%%%%%%%%%%%%%%%%%%%ù
\begin{table}[htb]
\centering
\caption{Comparison between the free energies obtained with the three models (BF, FK, and the present approach). 
Relative errors (in $\%$) are reported for different values of $\beta$ in the 
small- ($g=0.01$), intermediate- ($g=1$), and strong-coupling ($g=10$) regimes. $\omega=1$.}
\label{tab3}
\begin{tabular}{lcccc}
\hline
\hline
	&$\beta$&  $\;$ $\epsilon({\rm BF})(\%)$ $\;$ &  $\;$ $\epsilon({\rm FK})(\%)$ $\;$ & $\;$ $\epsilon(\rm This\;work)(\%)$ $\;$\\
\hline
$g=0.01$&       &       &         &\\
	&10     & 0.006 & 0.004   & -0.02   \\
	& 5     & 0.007 & 0.002   & -0.02   \\
	& 2     & 0.03  & 0.0002  & -0.03   \\
	& 1     & 0.9   & 0.0002  &  -0.3   \\
	& 0.1   & 0.5   & 0.01    &  -0.002   \\
\hline
$g=1$     &     &       &         &\\
        &10   & 1. & 1.      & -2.\\
        & 5   & 1. & 0.8     & -2.\\
        & 2   & 1. & 0.2     & -2.\\
        & 1   & 3. & 0.04    & -3.\\
        & 0.1 & 3. & 0.000001& -0.08\\
\hline
$g=10$    &     &       &         &\\
        &10   & 2. & 2.     & -4.\\
        & 5   & 2. & 2.     & -4. \\
        & 2   & 2. & 1.     & -4. \\
        & 1   & 2.2& 0.4    & -4. \\
        & 0.1 & 5. & 0.00006& -0.4\\
\hline
\hline
\end{tabular}
\end{table}
%%%%%%%%%%%%%%%%%%%%%%%%%%%%%%%%%%%%%%%%%%%%%%%%%%%%%%%%%%%%%%%%%%%%%%%%%%%%%%%%%%%%%%%%%%%%%%%%%%%%ùù
In Table \ref{tab3} a comparison between the free energies obtained with the three models (BF, FK, and our model) 
is presented.
Results are given at some selected values 
of $\beta$ and coupling constants $g$. For each approach, the relative errors $\epsilon$ (in $\%$) are reported.
A first remark is that the BF and FK free energies have positive errors, a result expected due to the variational character of both theories.
In constrast, as already noted above, our free energies
are found to be systematically smaller than the exact ones, a property which is {\it a priori} not expected.
Unfortunately, we have not been able to show whether or not this property is true, 
in particular when other types of potentials $V(x)$ are considered.
Now, taking this property for granted, it is tempting to maximize the energy with respect to the effective frequency, 
${\bar \omega}_g$, considered as a variational parameter [note 
that $E_0(g)$ and $E_1(g)$ depends only on ${\bar \omega}_g$].
Quite surprisingly, no improvement has been observed at all couplings, a result which would indicate that our {\it parameter-free} 
partition function is already optimal if the property is true. 
A second remark is that, as already noted by Srivastava and Vishwamittar\cite{Srivastava_1991}, the 
quality of the FK free energies is superior to that of the BF approach. This is particularly true in the high-temperature limit 
where the BF approach fails to converge to the classical limit. The present method, which has both the exact harmonic and classical limits, like in  
the FK approach, leads also to small errors over the full range of temperatures and couplings (here, maximum error of about $4\%$). 
However, the FK approach remains superior (maximum error of about $2\%$), particularly at high temperatures where 
the convergence of the FK partition function to the classsical limit is particularly rapid.\\

\begin{figure}[htb]
\centering
\includegraphics[scale=0.45]{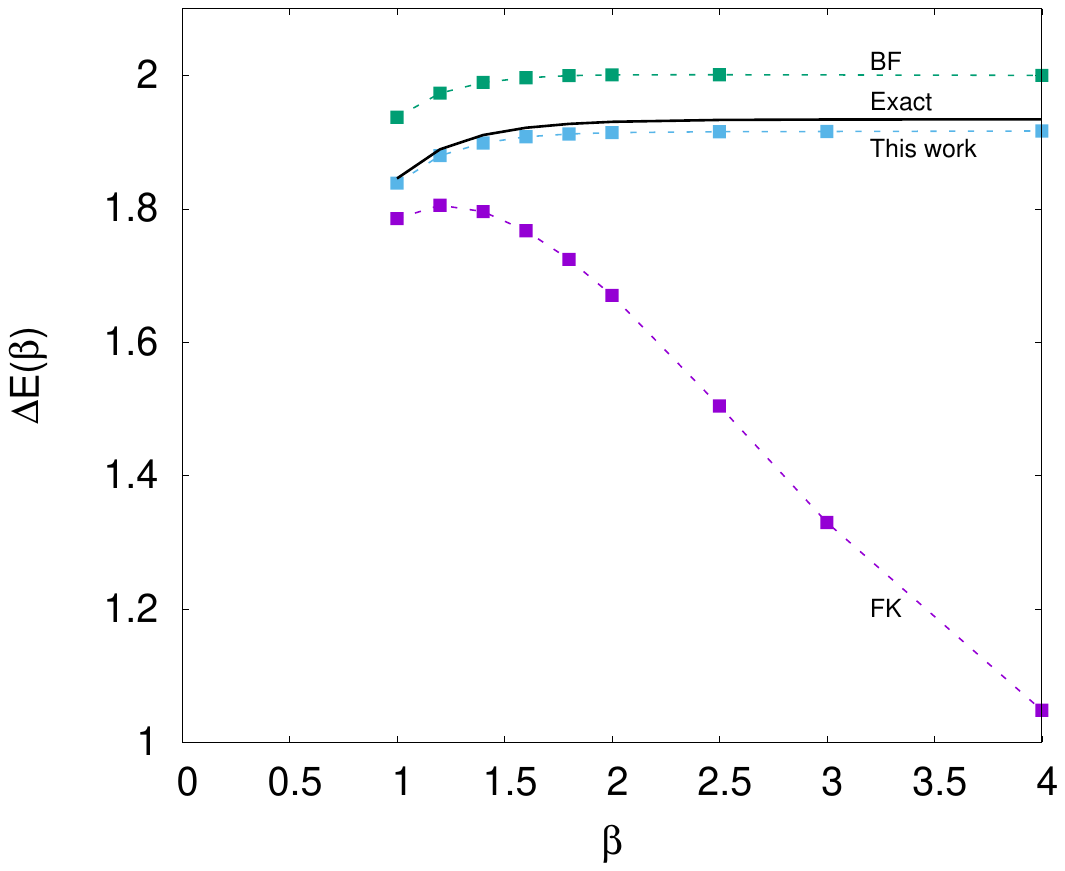}
\caption{Comparison between the temperature-dependent energy gaps as a function of $\beta$. The energy gap, $\Delta E=E_1(g)-E_0(g)$, is obtained 
as $\beta \rightarrow \infty$. The solid line gives the exact values. $\omega=1$ and $g=1$.}
\label{fig4}
\end{figure}

An important issue regarding approximate partition functions 
is their ability to reproduce the gap in energy, $\Delta E= E_1(g)-E_0(g)$, a critical quantity 
for the low-energy properties of the model.
In Fig.\ref{fig4} we present the temperature-dependent
energy gap defined as follows
\be
\Delta E(\beta) = E_1(\beta) -E_0(\beta)
\ee
with
\be
E_0(\beta)= -\frac{1}{\beta} \log{Z(\beta)}
\ee
and
\be
E_1(\beta)= -\frac{1}{\beta} \log{\qty[Z\qty(\beta)-e^{-\beta E_0\qty(\beta \rightarrow \infty)}]}.
\ee
At $\beta \rightarrow \infty$,
$\Delta E(\beta)$ is expected to converge to the energy gap. 
In Fig.\ref{fig4} the gap in energy as a function of $\beta$ is presented for $g=1$ (intermediate coupling).
As seen, because of its very construction, the FK approach is not at all suited for reproducing 
the energy gap. In contrast, this is not the case for the two other methods. In the case of our model, the
partition function has been modified by adding to it the small correction $\delta Z = e^{-\beta (E_0+{\bar \omega_g})} \qty[
	e^{-\beta \Delta} - (1-\beta \Delta)]$ in order to stabilize its zero-temperature limit
(see, the discussion above about the excited-state energies).
Both approaches converge 
to accurate values with an error of about $3.4\%$ for the BF approach and a significantly smaller error of about $0.9\%$ for the present approach. 

Finally, it is important to emphasize that, in this section, we have only compared {\it simple} formulas for the partition function. For the FK and BF 
methods the simple formulae used here correspond only to the lowest order of approximation in their formalism. 
In the case of FK, systematic corrections to the calculation of 
the effective classical potential can be performed,\cite{Kleinert_1993} thus leading to systematically better energies and gaps. 
It is also true in the variational framework for free energies, see for example\cite{Vlachos_1993}. 
However, in both cases, the formalism and formulas needed to evaluate the systematic corrections to the 
simple formulas become much more involved and the advantage of having a simple model is lost.

%%%%%% FIG4 %%%%%%%%%%%%%%%%%%%%%%%%%%%%%%%%%%%%%%%%%%%%%%%%%%%%%%%%%%%%%%%%%%%%%%%%%%%%%%%%%%%%%%
%%%%%%%%%%%%%%%%%%%%%%%%%%%%%%%%%%%%%%%%%%%%%%%%%%%%%%%%%%%%%%%%%%%%%%%%%%%%%%%%%%%%%%%%%%%%%%%%%%%%%
%%%%%%%%%%%%%%%%%%%%%%%%%%%%%%%%%
\section{Summary and perspectives}
%%%%%%%%%%%%%%%%%%%%%%%%%%%%%%%%%
\label{summa}
Let us first briefly summarize the derivation of our model partition function.
The first step is to express the partition function as a path integral, Eq.(\ref{Zex}).
Next, a gaussian approximation for the quartic contribution is introduced, Eq.(\ref{gauss_approximation}). 
This approximation amounts to reformulating $Z$
as the partition function of a harmonic oscillator with an effective frequency depending both on 
the temperature and coupling constant $g$,
Eq.(\ref{eqbeta}). However, because of the gaussian approximation 
the limit $n \rightarrow \infty$ for $Z_n$ is no longer defined, Eq.(\ref{noconvergence}). 
To address this problem, we introduce a Principle of Minimal Sensitivity 
of the PF to the effective frequency, Eq.(\ref{cond}). Thanks to this principle, 
we are able to propose our model partition function, Eq.(\ref{Zdef}) with Eqs.(\ref{Zdef_I},\ref{eq1},\ref{eq2}).

Despite its apparent simplicity, this partition function encapsulates a number of important properties which can be summarized as
folllows.\\
i) Both the harmonic ($g \rightarrow 0)$ and classical ($\beta \rightarrow 0$) limits are recovered.\\
ii) The free energy is accurate to a few percent over the entire range of temperatures and coupling strengths.\\
iii) The ground-state and first-excited energies are accurately reproduced for all coupling strengths, the energy gap 
having a typical error of about one percent.\\
iv) The divergence of the weak-coupling expansion
with a factorial-like growth of the coefficients is recovered. In addition, the coefficients are found to be rational numbers as the exact ones.\\
v.) The functional form of the strong-coupling expansion is reproduced. The leading coefficient 
is obtained with an error of a few percent.

The quartic oscillator being one of the simplest paradigmatic examples
of non-trivial quantum system, it has been and still will be the subject of numerous studies in quantum, statistical, and 
quantum field theory domains
to test new ideas and algorithms.
Then, we believe that to have at our disposal a simple analytic model for the partition function 
reproducing the key features of the exact partition function can be of 
interest to the community.
Note that the approach presented here can be generalized without difficulty to one-dimensional oscillators 
with an arbitrary anharmonicity (for example,
$x^p$ with $p$ even). Finally, it will be interesting to explore the applicability of the approach to multi-dimensional systems.
%%%%%%%%%%%%%%%%%%%%%%%%%%%%%%%%%%%%%
\acknowledgments{I would like to thank the Centre National de la Recherche Scientifique (CNRS) for its continued support.
I also acknowledge funding from the European Research Council (ERC) under the European Union's Horizon 2020 research and innovation programme (Grant agreement No.~863481).
}
%%%%%%%%%
\appendix
%%%%%%%%%

%%%%%%%%%%%%%%%%%%%%%%%%%%%%%%%%%%%%%%%%%%%%%%%%%%%%%%%%%%%%%%%%%%%%%%%%%%%%%%%%%%%%%%
\section{Derivation of $\prod_{k=1}^{n-1} \sin{ \frac{\pi k}{n}} =  \frac{n}{2^{n-1}}$}
%%%%%%%%%%%%%%%%%%%%%%%%%%%%%%%%%%%%%%%%%%%%%%%%%%%%%%%%%%%%%%%%%%%%%%%%%%%%%%%%%%%%%%%
\label{appendix_A}
The derivation of this equality can be found at several places in the literature. However, for completeness, we provide the following derivation.\\

For $1 \le k \le n-1$ we have
\be
| 1 -e^{i 2\pi \frac{k}{n}}| =2 \sin{\frac{\pi k}{n}}.
\ee
The product of sine can thus be written as
\be
\prod_{k=1}^{n-1} \sin{ \frac{\pi k}{n}} = \frac{1}{2^{n-1}} \prod_{k=1}^{n-1} | 1 -e^{2\pi i\frac{k}{n}}|
= \frac{1}{2^{n-1}} | \prod_{k=1}^{n-1} (1 -e^{2\pi i\frac{k}{n}})|.
\ee
Let us introduce the roots of the unity,
$w_k= e^{i \frac{2\pi k}{n}}$, solutions of $x^n -1 =0$.
We have
\be
x^n -1 =\prod_{k=0}^{n-1} (x - w_k).
\ee
Defining $Q(x)$ as
\be
Q(x)=\prod_{k=1}^{n-1} (x - w_k)
\ee
and writing
\be
x^n-1 = (x-1) (1 + x + ... + x^{n-1})
\ee
we then get
\be
Q(x)= 1 + x + ... + x^{n-1}
\ee
and
\be
|\prod_{k=1}^{n-1} (1 -e^{2\pi i\frac{k}{n}})| = |Q(1)|=n.
\ee
Finally,
\be
\prod_{k=1}^{n-1} \sin{ \frac{\pi k}{n}} =  \frac{n}{2^{n-1}}.
\ee
%%%%%%%%%%%%%%%%%%%%%%%%%%%%%%%%%%%%%%%%%%%%%%%%%%%%%%%%%%%%%
\section{Derivation  of $\lim_{n \rightarrow \infty} P_n(x)$}
%%%%%%%%%%%%%%%%%%%%%%%%%%%%%%%%%%%%%%%%%%%%%%%%%%%%%%%%%%%%%
\label{appendix_B}
$P_n(x)$ is defined as
\be
P_n(x)= \prod_{k=2}^n  \frac{1}{ \sqrt{1 + \frac{x^2}{2\lambda_k n^2}}}
\label{pn}
\ee
where 
\be
\lambda_k = 1 - \cos{ \frac{2\pi}{n}(k-1)}   \;\;\; k=1,n.
\ee
Introducing the logarithm of $P_n$
\be
\ln{ P_n} = -\frac{1}{2} \sum_{k=2}^n \ln{\qty( 1 + \frac{x^2}{2\lambda_k n^2})}
\ee
and the Taylor expansion of $\ln{\qty(1+x)}$
\be
\ln{\qty( 1+x)} = \sum_{l=1}^{+\infty} \frac{(-1)^{l-1}}{l}x^l
\ee
we get
\be
\ln{ P_n} = \frac{1}{2} \sum_{l=1}^{+\infty} \frac{(-1)^{l}}{l} \qty(\frac{x^2}{2 n^2})^l S^l_n
\label{lnpn}
\ee
where
\be
S^l_n= \sum_{k=2}^n \qty( \frac{1}{\lambda_k})^l = \sum_{k=2}^n {\qty( \frac{1}{1-\cos{\qty(2\pi \frac{k-1}{n})}})}^l.
\ee
$S^l_n$ can be rewritten as
\be
S^l_n= \frac{1}{2l} \sum_{k=1}^{n-1} \frac{1}{ \sin^{2l}\qty(\pi \frac{k}{n})}.
\ee
As shown  by Fisher\cite{Gardner_1971}, the large-$n$ behavior of $S^l_n$ can be related to the 
Riemann zeta function as follows
\be
\lim_{n \rightarrow \infty} n^{-2l} \sum_{k=1}^{n-1} \frac{1}{\sin^{2l}{\qty(\pi \frac{k}{n})}} 
= 2 \frac{\zeta(2l)}{\pi^{2l}}.
\ee
Now, using the well-known equality 
\be
\zeta(2l) =(-1)^{l+1} \frac{B_{2l} (2 \pi)^{2l}}{2 (2l)!} 
\ee
we get the asymptotic behavior at large $n$ of $S^l_n$
\be
S^l_n \sim_{n \rightarrow \infty} \frac{ |B_{2l}| 2^l}{(2l)!} n^{2l}
\label{snl}
\ee
where $B_{2l}$ are the Bernoulli numbers. From now on, all following expressions must be understood as 
valid only in the large-$n$ regime.
Using Eqs.(\ref{lnpn}) and (\ref{snl}) we get
\be
\ln{ P_n} = \frac{1}{2} \sum_{l=1}^{+\infty} (-1)^l x^{2l} \frac{ |B_{2l}|}{l (2l)!}.
\label{pn2}
\ee
Let us define
\be
u(x)= \frac{1}{2} \sum_{l=1}^{+\infty} (-1)^l x^{2l} \frac{ |B_{2l}|}{l (2l)!}
\ee
so that
\be
P_n= e^{u(x)}.
\label{pnu}
\ee
For $l \ge 1$ we have
\be
|B_{2l}|= (-1)^{l-1} B_{2l}.
\ee
Then, $u(x)$ can be written as
\be
u(x)= -\frac{1}{2} \sum_{l=1}^{+\infty} x^{2l} \frac{ B_{2l}}{l (2l)!}.
\ee
Now, taking the derivative of $u(x)$
\be
u'(x)= -\frac{1}{x} \sum_{l=1}^{+\infty} x^{2l} \frac{ B_{2l}}{(2l)!}
\ee
and using the fact that
\be
\coth{x} = \sum_{n=0}^{+\infty} \frac{2^{2n} B_{2n}}{(2n)!} x^{2n-1}
= \frac{1}{x} \sum_{n=0}^{+\infty} \frac{B_{2n}}{(2n)!} (2x)^{2n}
\ee
the derivative of $u(x)$ is found to be
\be
u'(x)= -\frac{1}{2} \coth{\frac{x}{2}} + \frac{1}{x}.
\ee
Using Eq.(\ref{pnu}) and $P_n(0)=1$, we can write
\be
P_n(x)= e^{ \int_0^x dy u'(y)}.
\ee
Finally, after integrating we get 
\be
P_n(x)= \frac{x}{e^{-\frac{x}{2}} - e^{-\frac{x}{2}}}.
\ee
%%%%%%%%%%%%%%%%%%%%%%%%%
\section{Strong coupling}
%%%%%%%%%%%%%%%%%%%%%%%%%
\label{appendix_C}
The function $B(x)$ is defined as
\be
B(x) = \frac{1}{2} \frac{\int dy \; e^{-y^2 - x y^4}} { \int dy \; y^2 e^{-y^2 - x y^4}}  \;\;\; {\rm for} \; x \ge 0.
\ee
After a change of variable, it can be rewritten as
\be
B(x) =  \frac{\sqrt{x}}{2} \frac{\int dz \; e^{-\frac{1}{\sqrt{x}} z^2 - z^4}} {\int dz \; z^2 e^{-\frac{1}{\sqrt{x}} z^2 - z^4}}.
\ee
The strong coupling expansion of $B(x)$ at large $x$ has thus the following form
\be
B(x) =  \sqrt{x} \sum_{n=0}^{+\infty} B_n \qty(\frac{1}{\sqrt{x}})^n.
\ee
The three first coefficients are given by
\be
B_0= 2 \frac{ \Gamma \qty(\frac{5}{4})} {\Gamma \qty(\frac{3}{4})}
\ee
\be
B_1= \frac{ 4\Gamma^2\qty(\frac{5}{4}) - \Gamma^2\qty(\frac{3}{4}) }{2 \Gamma^2\qty(\frac{3}{4})}
\ee
and
\be
B_2= -\frac{ \Gamma \qty(\frac{5}{4}) \qty[ \Gamma^2\qty(\frac{3}{4}) - 8 \Gamma^2\qty(\frac{5}{4}) + 4 \Gamma\qty(\frac{3}{4}) \Gamma\qty(\frac{7}{4}) ] }{4 \Gamma^3\qty(\frac{3}{4})}.
\ee
As shown in the text, the expansion of ${\bar \omega_g}(g)$ writes
\be
{\bar \omega_g}(g)= (2g)^{\frac{1}{3} }\sum_{n=0}^{+\infty} \omega_n y^n 
\ee
where the variable $y$ is defined as
\be
y = g^{-\frac{2}{3}}.
\ee
The coefficients $\omega_n$ are expressed in terms of the coefficients $B_p$'s ($p \le n$) as follows
\be
\omega_0= {B_0}^{\frac{2}{3}}
\ee
\be
\omega_1= \frac{ 2^{\frac{2}{3}} \omega B_1}{ 3 {B_0}^{\frac{2}{3}}}
\ee
\be
\omega_2= \frac{\omega^2}{2^{\frac{2}{3}}} \qty(  \frac{2B_0 B_2 -{\qty(B_1)}^2}{ 3 {\qty(B_0)}^2})  \;\; {\rm etc.}
\ee
The expression of the ground-state energy as a function of ${\bar \omega_g}$ writes
\be
E_0(g) = \frac{ {\bar \omega_g} }{2} \qty[ 1 -\ln{ \qty(\sqrt{\frac{\bar \omega_g}{\pi}} {\bar I}(g))} ]
\label{E0_final2}
\ee
where
\be
{\bar I}(g)= \int dx e^{ -\frac{2}{{\bar \omega_g}} \qty(\frac{1}{2} \omega^2 x^2 + g x^4)}.
\ee
Defining $J(g)$ as
\be
J(g) = \sqrt{\frac{\bar \omega_g}{\pi}} {\bar I}(g) = 
\sum_{n=0}^{+\infty} J_n y^n
\ee
the three first coefficients $J_n$ are found to be
\be
J_0= \frac{ \omega^{\frac{3}{4}}_0 \Gamma(\frac{1}{4}) }{ 2 \sqrt{\pi}}
\ee
\be
J_1= \frac{ 2^{\frac{1}{3}} \omega^2 \sqrt{\omega_0} \Gamma\qty(-\frac{1}{4}) + 6  \omega_1 \Gamma(\frac{1}{4}) }
{ 16 \sqrt{\pi} {\omega_0}^{\frac{1}{4}}}
\ee
and
\be
J_2= \frac{ 2^{\frac{1}{3}}  \sqrt{\omega_0} \omega_1 \omega^2 \Gamma\qty(-\frac{1}{4}) + \qty( 2^{\frac{2}{3}} \omega^4 \omega_0 
- 3\omega^2_1 + 24 \omega_0 \omega_2 ) \Gamma(\frac{1}{4}) }
{ 64 \sqrt{\pi} {\omega_0}^{\frac{5}{4}}}.
\ee
Writing $E_0(g)$ under the form
\be
E_0(g)= \frac{ \qty(2g)^{\frac{1}{3}} }{2}  \qty(\sum_{n=0}^{+\infty} \omega_n y^n) \qty[ 1 -\ln{ \qty(
\sum_{n=0}^{+\infty} J_n y^n )} ]
\ee
we have
\be
E_0(y)= g^{\frac{1}{3}} 
\sum_{n=0}^{+\infty} \alpha_n  y^n.
\ee
The first coefficients are given by
\be
\alpha_0= \omega_0 ( 1 - \ln{J_0})
\ee
\be
\alpha_1= -\frac{J_1 \omega_0}{J_0}+ \omega_1 ( 1 - \ln{J_0})
\ee
and
\be
\alpha_2= \frac{(J^2_1 -2 J_0 J_2) \omega_0} {2 J^2_0} - \frac{ J_1 \omega_1}{J_0}+ \omega_2 ( 1 - \ln{J_0}).
\ee
%%%%%%%%%%%%%%%%%%%%%%%%%%%%%%%%%%%%%%%%%%%%%%%
\section{Feynman-Kleinert variational approach}
%%%%%%%%%%%%%%%%%%%%%%%%%%%%%%%%%%%%%%%%%%%%%%%
\label{appendix_D}
Using the notations of \cite{Feynman_1986} (in particular, $t \equiv \beta$), the basic equations of the Feynman-Kleinert variational approach are
\be
Z= \frac{1}{\sqrt{2\pi t}} \int dx_0 e^{-t W_1(x_0)}
\ee
where $W_1$ is an effective classical potential obtained as
\be
W_1(x_0) = \min_{a^2,\Omega} {\tilde W}_1(x_0,a^2,\Omega)
\ee
and ${\tilde W}_1$ is given by
\be
{\tilde W}_1(x_0,a^2,\Omega) = \frac{1}{t} \ln{ \frac{ \sinh{\frac{\Omega t} {2}}}{\frac{\Omega t} {2} }}-\frac{\Omega^2}{2} a^2 + V_{a^2}(x_0).
\ee
Here, the potential $V_{a^2}$ is defined by
\be
V_{a^2}(x_0) = \int \frac{ dx^\prime }{\sqrt{2 \pi a^2}} e^{ - \frac{1}{2 a^2} (x_0 - x^\prime)^2 } V(x^\prime).
\ee
After minimization of ${\tilde W}_1$ a pair of optimal parameters depending on $x_0$, ($a^2(x_0),\Omega(x_0))$ is obtained.
The minimization of ${\tilde W}_1$ with respect to $\Omega$ gives the equation
\be
a^2= \frac{1}{\Omega^2 t} \qty( \frac{\Omega t }{2} \coth{\frac{\Omega t}{2}} -1 )
\ee
and the minimization with respect to $a^2$ gives
\be
\Omega^2 =  \frac{ \partial^2 V_{a^2}(x_0)} {\partial x_0^2}.
\ee
Let us explicit these equations in the case of the quartic oscillor,
$V(x) = \frac{1}{2} \omega^2 x^2 + g x^4$. We get
\be
V_{a^2}(x) = \frac{1}{2} a^2 \omega^2  + 3 a^4 g + \frac{1}{2} \qty( 12 a^2 g + \omega^2) x^2 + g x^4
\ee
and
\be
\Omega^2(x)  = 12 a^2 g + \omega^2 + 12 g x^2.
\ee
Now, to get the ground-state energy, large-time limit has to be considered.
As $t \rightarrow \infty$ we have
\be
\frac{1}{t} \ln{ \frac{ \sinh{\frac{\Omega(x_0) t} {2}}}{\frac{\Omega(x_0) t} {2} }} \rightarrow \frac{\Omega(x_0)}{2}
\ee
and then
\be
a^2(x_0)=\frac{1}{2\Omega(x_0)}.
\ee
The equation determining $\Omega(x_0)$ is
\be
\Omega(x_0)^3 -\Omega(x_0) \qty(\omega^2 + 12 g x^2_0) -6g =0
\label{p3}
\ee
and the partition function writes
\be
Z= \frac{1}{\sqrt{2\pi t}} \int dx_0 e^{-t f(x_0)}
\ee
with
\be
f(x_0)= \frac{\Omega(x_0)}{4} + V_{a^2(x_0)}(x_0).
\ee
Now, in the large-time limit only the constant part of $f(x)$ will give a contribution to $-\frac{1}{t} \ln{Z}$, we then have
\be
E_0(g) = f(0).
\ee
\\
The real solution of the third-order polynomial equation, Eq.(\ref{p3}), with $x_0=0$ is
\be
\Omega_0(g)= \frac{\omega^2}{ 3^{\frac{1}{3} } \Delta(g)^{\frac{1}{3}}} + \frac{ \Delta(g)^{\frac{1}{3}}} { 3^{\frac{2}{3}}}
\ee
where 
\be
\Delta(g)= 27 g + \sqrt{3} \sqrt{ 243 g^2 - \omega^6}.
\ee
Finally, the ground-state energy of the Feynman-Kleinert model writes
\be
E_0(g)= \frac{\Omega_0(g)}{4} + \frac{\omega^2}{4\Omega_0(g)} + \frac{3g}{4\Omega^2_0(g)}.
\ee
%%%%%%%%%%%%%%%%%%%%%%%%%%%%%%%%%%%%%%%%%%%%%%%%%%%
\section{B\"uttner-Flytzanis  variational approach}
%%%%%%%%%%%%%%%%%%%%%%%%%%%%%%%%%%%%%%%%%%%%%%%%%%%
\label{appendix_E}
In the B\"uttner-Flytzanis variational approach,\cite{Buttner_1987} the free energy is written as 
\be
F(\beta)=F_0(\beta) + \frac{\omega^2-{\tilde \omega}^2}{4} 
\qty[  \frac{ \coth{ \frac{{\tilde \omega \beta}}{2}}}{ {\tilde \omega}}]
+ \frac{3g}{4} \qty[  \frac{ \coth{ \frac{{\tilde \omega \beta}}{2}}}{ {\tilde \omega}}]^2
\label{Fb}
\ee
where the frequency $ {\tilde \omega}$ is the variational parameter and
$F_0(\beta)$ the free energy of the harmonic oscillator with frequency ${\tilde \omega}$
\be
F_0(\beta) = \beta^{-1} \ln \qty[2 \sinh{\qty( \frac{ {\tilde \omega} \beta}{2})} ].
\ee
Minimizing Eq.(\ref{Fb}) with respect to ${\tilde \omega}$ gives the equation determining ${\tilde \omega}$
\be
{\tilde \omega}^3 - \omega^2 {\tilde \omega} -6g \coth{\qty( \frac{ {\tilde \omega} \beta}{2})}=0.
\ee
%%%%%%%%%%%%%%%%%%%%%%%%%%%%%%%%%%%%%%%%%%%%%%%%%%%
%\bibliography{paper}

%%%%%%%%%%%%%%%%%%%%%%%%%%%%%%%%%%%%%%%%%%%%%%%%%%%
\end{document}